\documentstyle[preprint,revtex]{aps}
\begin{document}
\draft
\preprint{IFT-P.008/92}
\hfill{hep-ph/9206242}
\begin{title}
$SU(3)\otimes U(1)$ Model for Electroweak Interactions
\end{title}
\author{ F. Pisano and V. Pleitez}
\begin{instit}
Instituto de F\'\i sica Te\'orica \\
Universidade Estadual Paulista \\
Rua Pamplona, 145 \\
CEP 01405--S\~ao Paulo, SP \\
Brazil
\end{instit}
\begin{abstract}
We consider a gauge model based on a $SU(3)\otimes U(1)$ symmetry in
which the lepton number is violated explicitly by charged scalar and
gauge bosons, including a vector field with double electric charge.
\end{abstract}
\pacs{PACS numbers:}
\section{Introduction}
\label{sec:intro}
Some years ago, it was pointed out that processes like
$e^-e^-\rightarrow W^-V^-$ in Fig.\ \ref{fig1}(a),
if induced by right-handed currents coupled to the vector $V^-$,
imply violation of unitarity at high energies. Then, if the
right-handed currents are part of a gauge theory, it has been argued
that at least some neutrinos must have non-zero mass~\cite{bk}.

The argument to justify this follows exactly the same way as in the
usual electroweak theory for the process $\nu\bar\nu\to
 W^+W^-$. The graph induced by an electron exchange has bad
high-energy behavior; when the energy goes to infinity, the
respective amplitude violates unitarity~\cite{q}.

In Fig.\ \ref{fig1}(a) the lower vertex indicates a right-handed
current which absorbs the right-handed antineutrino coming from the
upper vertex, which represents the left-handed current of the
electroweak standard model. The part of the amplitude, corresponding
to Fig.\ \ref{fig1}(a), in which we are interested is
\begin{equation}
\sum_{\nu_m}U^L_{em}\frac{\slash\!\!\!q}{q^2-M^2_{\nu_m}}U^R_{em}\;,
\label{e1}
\end{equation}
where $U^L(U^R)$ is the mixing matrix in the left(right)-handed
current, and $q$ is the 4-momentum transfer~\cite{bk}. The
space-time structure of Eq.~(\ref{e1}) is the same as the charged
lepton exchange amplitude in the process $\nu\bar\nu\to
W^+W^-$~\cite{q}. Then, we must  have the same bad high energy
behavior of the last process. One way to avoid this is to have
cancellation among the contributions from the various $\nu_m$
exchanges when we add them up; at high energy and large $q^2$, the
latter dominates the denominator in Eq.~(\ref{e1}), and if we require
that
\begin{equation}
\sum_{\nu_m}U^L_{em}U^R_{em}=0,
\label{e2}
\end{equation}
the amplitude in Eq.~(\ref{e1}) vanishes even at low energies,
unless at least one of the masses $M_{\nu_m}$ is non-zero. On the
other hand, the diagram in Fig.\ \ref{fig1}(a) or its time reversed one,
$W^-W^-\to e^-e^-$, appearing in Fig.\ \ref{fig1}(b), when both
vertices are left-handed, proceeds via Majorana massive neutrinos.

Here we are concerned with a gauge model based on a $SU_L(3)\otimes
U_N(1)$ symmetry. The original motivation leading to the study of
this model stemmed from the observation that a gauge theory must be
consistent, that is, unitary and renormalizable, independently of the
values of some parameters, like mixing angles. Then, from
this point of view, instead of the condition in Eq.~(\ref{e2}) in
order to solve the problem placed by the graph in Fig.\
\ref{fig1}(a), we prefer the introduction of a doubly charged gauge
boson which, like the $Z^0$ in the standard model, will restore the
good high energy behavior.

Although there exist in the literature several models based on a
$SU(3)\otimes U(1)$ gauge symmetry \cite{v,su,ls,fm,lw}, our model
has a different representation content and a quite different new
physics at an, in principle, arbitrary mass scale. The main new
features of our model occur in processes in which the initial
electric charge is not zero. Even from the theoretical point of view,
that sort of processes have not been well studied, for instance,
general results exist only for zero initial charge~\cite{clt}.

The plan of this paper is as follows: Sec.~\ref{sec:model} is devoted
to present the model. Some phenomenological consequences are given in
Sec.~\ref{sec:pc}. In this way we can estimate the allowed value
for the mass scale characterizing the new physics. In Sec.~\ref{sec:sp} we
study briefly the scalar potential and show that there is not mixing
between the lepton-number-conserving and lepton-number-violating scalar
fields which could induce decays like the neutrinoless double beta
decay. The last section is devoted to our conclusions and some
comments and, in the Appendix we give more details about the
definition of the charge conjugation operation we have used in this
work.
\section{The Model}
\label{sec:model}
As we said before, the gauge model that we shall consider is one
in which the gauge group is $SU_L(3)\otimes U_N(1)$. This is possibly
the simplest way to enlarge the gauge group $SU_L(2)\otimes U_Y(1)$ in
order to have doubly charged gauge bosons, without losing the natural
features of the standard electroweak model. The price we must pay is
the introduction of exotic quarks, with electric charge $5/3$
and $-4/3$.

In this model we have the processes appearing in Figs.~\
\ref{fig2}(a) and  \ \ref{fig2}(b),the last diagram plays the same
role as the similar diagram with $Z^0$ in the standard model and it
restores the ``safe'' high energy behavior of the model. Both vector
bosons $V^-$ and $U^{--}$ in Figs.\ \ref{fig2}(a,b)  are very
massive, and their masses depend on the mass scale of the breaking of
the $SU_L(3)\otimes U_N(1)$ symmetry into $SU_L(2)\otimes U_Y(1)$.
Phenomenological bounds on this mass scale will be given in the next
section.
\subsection{Yukawa Interactions}
\label{subsec:yukawa}
We start by choosing the following triplet representations for the
left-handed fields of the first family,
\begin{equation}
\begin{array}{cccc}
E_L=\left(\begin{array}{c}
\nu_e \\  e\\ e^c
\end{array}\right)_L & (\underline{3},0);&
Q_{1L}=\left(\begin{array}{c}
 u \\ d \\ J_1
\end{array}\right)_L & (\underline{3},+\frac{2}{3}),
\end{array}
\label{e3}
\end{equation}
and
\begin{equation}
\begin{array}{ccc}
u_R \;(\underline{1},+\frac{2}{3}); & d_R
\;(\underline{1},-\frac{1}{3}); & J_{1R}
\;(\underline{1},+\frac{5}{3}),
\end{array}
\label{e4}
\end{equation}
for the respective right-handed fields. Notice that we have not
introduced right-handed neutrinos. The numbers $0,2/3$ in
Eq.~(\ref{e3}) and $2/3,-1/3$ and $5/3$ in Eq.~(\ref{e4}) are
$U_N(1)$ charges. The electric charge operator has been defined as
\begin{equation}
{Q\over e}= \frac{1}{2}\left(\lambda_3 -\sqrt{3}\lambda_8\right)+N,
\label{e5}
\end{equation}
where $\lambda_3$ and $\lambda_8$ are the usual Gell-Mann matrices;
$N$ is proportional to the unit matrix. Then, the exotic quark
$J_1$, has electric charge $+5/3$.

The other two lepton generations also belong to triplet
representations,
\begin{equation}
\begin{array}{cccc}
M_L=\left(\begin{array}{c}
\nu_\mu \\ \mu \\ \mu^c
\end{array}\right)_L & (\underline{3},0);&
T_L=\left(\begin{array}{c}
\nu_\tau \\ \tau \\ \tau^c
\end{array}\right)_L & (\underline{3},0).
\end{array}
\label{e6}
\end{equation}
The model is anomaly free if we have equal number of triplets and
antitriplets, counting the color of $SU(3)_c$, and furthermore
requiring the sum of all fermion charges to vanish. As in the model
of Ref.~\cite{v}, the anomaly cancellation occurs for the three
generations together and not generation by generation.

Then, we must introduce the antitriplets:
\begin{equation}
\begin{array}{cccc}
Q_{2L}=\left(\begin{array}{c}
J_2\\ c \\ s
\end{array}\right)_L & (\underline{3}^*,-\frac{1}{3});&
Q_{3L}=\left(\begin{array}{c}
J_3 \\ t \\ b
\end{array}\right)_L & (\underline{3}^*,-\frac{1}{3}),
\end{array}
\label{e7}
\end{equation}
also with the respective right-handed fields in singlets. The quarks
$J_2$ and $J_3$ have both charge $-4/3$.

In order to generate fermion masses, we introduce the following
Higgs triplets, $\eta, \rho$ and $\chi$:
\begin{equation}
\begin{array}{cccccc}
\left(\begin{array}{c}
\eta^0 \\ \eta^-_1 \\ \eta^+_2
\end{array}\right)& (\underline{3},0);&
\left(\begin{array}{c}
\rho^+ \\ \rho^0 \\ \rho^{++}
\end{array}\right) & (\underline{3},1);&
\left(\begin{array}{c}
\chi^- \\ \chi^{--} \\ \chi^0\end{array}\right) & (\underline{3},-1),
\end{array}
\label{e8}
\end{equation}
These Higgs triplets will produce the following hierarchical symmetry
breaking
\begin{equation}
SU_L(3)\otimes U_N(1)\stackrel{<\chi>}{\longrightarrow}SU_L(2)\otimes
U_Y(1)\stackrel{<\rho,\eta>}{\longrightarrow}U_{e.m}(1),
\label{e9}
\end{equation}
The Yukawa Lagrangian, without considering the mixed terms between
quarks, is
\begin{eqnarray}
-{\cal L}_Y&=&\frac{1}{2}\sum_lG_l
\varepsilon^{ijk}\bar\psi^c_{li}\psi_{lj}\eta_k+\bar Q_{1L}(G_uu_R\eta
+G_dd_R\rho+G_{J1}J_{1R}\chi) \nonumber \\
& &\mbox{}+\left(G_c\bar Q_{2L} c_R +G_t\bar Q_{3L} t_R\right)\rho^* +
\left(G_s\bar Q_{2L} s_R+G_b\bar Q_{3L} b_R\right)\eta^* \nonumber \\
& &\mbox{}+\left(G_{J2}\bar Q_{2L}J_{2R}+G_{J3}\bar
Q_{3L}J_{3R}\right)\chi^*+h.c.
\label{e10}
\end{eqnarray}
with $l=e,\mu,\tau$. Explicitly, we have for the leptons
\begin{equation}
2{\cal L}_{lY}\!=\!\sum_lG_l\left[(\bar l^c_Rl^c_L\!\!-\!\!\bar
l_Rl_L)\eta^0-(\bar \nu^c_{lR}l^c_L\!\!-\!\!\bar
l_R\nu_{lL})\eta^-_1+(\bar\nu^c_{lR}l_L\!\!-\!\!\bar
l^c_R\nu_{lL})\eta^+_2\right]+h.c. ,
\label{e11}
\end{equation}
and using the definition of charge conjugation,
$\psi^c=\gamma^5C\bar\psi^T$, that we shall discuss in the Appendix, we can
write Eq.~(\ref{e11}) as
\begin{equation}
{\cal L}_{lY}=\sum_lG_l(-\bar l_Rl_L\eta^0+\bar l_R\nu_L\eta^-_1 +
\bar\nu^c_Rl_L\eta^+_2+h.c.).
\label{e12}
\end{equation}
In Eq.~(\ref{e12}) there is lepton number violation through the
coupling with the $\eta_2^+$ Higgs scalar.

For the first and second quark generations we have the following
Yukawa interactions
\begin{eqnarray}
-{\cal L}_{QY}&=& G_u(\bar u_Lu_R\eta^0+\bar d_Lu_R\eta^-_1+\bar
J_{1L}u_R\eta^+_2) \nonumber \\ & &\mbox{} +G_d(\bar u_Ld_R\rho^++\bar
d_Ld_R\rho^0+\bar J_{1L}d_R\rho^{++}) \nonumber \\ & &\mbox{}
+ G_c(\bar J_{2L}c_R\rho^{--}+\bar c_Lc_R\rho^{0*}+\bar
s_Lc_R\rho^-) \nonumber \\ & &\mbox{}
+G_s(\bar J_{2L}s_R\eta_2^-+\bar c_Ls_R\eta_1^++\bar
s_Ls_R\eta^{0*}) \nonumber \\ & &\mbox{}
+G_{J_1}(\bar u_LJ_{1R}\chi^-+\bar
d_LJ_{1R}\chi^{--}+\bar J_{1L}J_{1R}\chi^{0*})\nonumber \\
& &\mbox{}+G_{J_2}(\bar J_{2L}J_{2R}\chi^0+\bar c_LJ_{2R}\chi^{++}+\bar
s_LJ_{2R}\chi^+)+h.c.
\label{e13}
\end{eqnarray}
The Yukawa interactions for the third quark generation is obtained
from those of the second generation making $c\to t$,
$s\to b$ and $J_2\to J_3$.  In Eq.~(\ref{e10}), since
the neutrinos are massless there is not mixing between leptons,
it is not necessary at all to consider terms like
$\frac{1}{2}\sum_{n,m}h_{lm}\bar\psi^c_{niL}\psi_{mjL}H^{[ij]}+h.c.$
where $n,m=e,\mu,\tau$, the coupling constants $h_{nm}=-h_{mn}$ and
$H^{[ij]}=\varepsilon^{ijk}\eta^k$.

The neutral component of the Higgs fields develops
the following vacuum-expectation value,
\begin{equation}
\begin{array}{cccccc}
<\eta^0>=&\frac{1}{\sqrt2}\left(\begin{array}{c}
v_\eta \\ 0 \\ 0
\end{array}\right), &
<\rho^0>=&\frac{1}{\sqrt2}\left(\begin{array}{c}
0 \\ v_\rho \\ 0
\end{array}\right), &
<\chi^0>=&\frac{1}{\sqrt2}\left(\begin{array}{c}
0 \\ 0 \\ v_\chi\end{array}\right).
\end{array}
\label{e14}
\end{equation}
So, the masses of the fermions are
$m_l=G_l\frac{v_\eta}{\sqrt2}$, for the charged leptons and
\begin{equation}
\begin{array}{cclccc}
m_u=&G_u\frac{v_\eta}{\sqrt2},&m_c=&G_c\frac{v_\rho}{\sqrt2},&
m_t=&G_t\frac{v_\rho}{\sqrt2},\\
m_d=&G_d\frac{v_\rho}{\sqrt2},&m_s=&G_s\frac{v_\eta}{\sqrt2},&
m_b=&G_b\frac{v_\eta}{\sqrt2},\\
m_{J_1}=&G_{J_1}\frac{v_\chi}{\sqrt2},&m_{J_2}=&G_{J_2}\frac{v_\chi}{\sqrt2},&
m_{J_3}=&G_{J_3}\frac{v_\chi}{\sqrt2},
\end{array}
\label{e15}
\end{equation}
for the quarks. The exotic quarks obtain their masses from the
$\chi$-triplet. Notice that, if we had had introduced right-handed
neutrinos, we would have massive Dirac neutrinos through their
couplings with the $\eta$ Higgs triplet.
\subsection{The Gauge Bosons}
\label{gb}
The gauge bosons of this theory consist of an octet $W^a_\mu$
associated with $SU_L(3)$ and a singlet $B_\mu$ associated with
$U_N(1)$. The covariant derivatives are:
\begin{equation}
{\cal D}_\mu\varphi_i=\partial_\mu\varphi_i+ig(\vec
W_\mu\cdot\frac{\vec\lambda}{2})_i^j\varphi_j
+ig'N_\varphi\varphi_i B_\mu,
\label{e16}
\end{equation}
where $N_\varphi$ denotes the $N$ charge for the $\varphi$ Higgs
multiplet, $\varphi=\eta,\rho,\chi$. Using Eqs.~(\ref{e14})
in Eq.~(\ref{e16}) we obtain the symmetry breaking pattern appearing
in Eq.~(\ref{e9}).

The gauge bosons $\sqrt2W^+\equiv -(W^1\!-\!iW^2)$, $\sqrt2V^-\equiv
-(W^4\!-\!iW^5)$ and $\sqrt2U^{--}\equiv -(W^6\!-\!iW^7)$ have the
following masses:
\begin{equation}
M^2_W=\frac{1}{4}g^2\left(v^2_\eta+v^2_\rho\right)\,;
M^2_V=\frac{1}{4}g^2\left(v^2_\eta+v^2_\chi\right)\,;
M^2_U=\frac{1}{4}g^2\left(v^2_\rho+v^2_\chi\right).
\label{e17}
\end{equation}
Notice that even if $v_\eta=v_\rho\approx v/\sqrt2$, being $v$ the
usual vacuum expectation value of the Higgs in the standard model,
the $v_\chi$ must be large enough in order to keep the new gauge
bosons, $V^+$ and $U^{++}$, sufficiently heavy in order to have
consistence with low energy phenomenology. On the other hand, the
neutral gauge bosons have the following mass matrix in the
$(W^3,W^8,B)$ basis
\begin{equation}
M^2=\frac{1}{4}g^2\left(\begin{array}{ccc}
v^2_\eta+v^2_\rho & \frac{1}{\sqrt3}(v^2_\eta-v^2_\rho) &
-2\frac{g'}{g}v^2_\rho \\
\frac{1}{\sqrt3}(v^2_\eta-v^2_\rho)&
\frac{1}{3}(v^2_\eta+v^2_\rho+4v^2_\chi)
 & \frac{2}{\sqrt3}\frac{g'}{g}(v^2_\rho+2v^2_\chi) \\
-2\frac{g'}{g}v^2_\rho &
\frac{2}{\sqrt3}\frac{g'}{g}(v^2_\rho+2v^2_\chi) &
4\frac{g'^2}{g^2}(v^2_\rho+v^2_\chi)
\end{array}\right),
\label{e18}
\end{equation}
and, since $\det M^2=0$ we must have a photon after the symmetry
breaking. If we had had introduced a $\underline{6}^*$, the matrix
$M^2$ in Eq.~(\ref{e18}) would be such that $\det M^2\not=0$.
In fact, the eingenvalues of the matrix in Eq.~(\ref{e18}) are:
\begin{equation}
M^2_A=0,\qquad
M^2_{Z}\simeq
\frac{g^2}{4}\frac{g^2+4g'^2}{g^2+3g'^2}(v^2_\eta+v^2_\rho),\qquad
M^2_{Z'}\simeq\frac{1}{3}(g^2+3g'^2)v^2_\chi,
\label{e19}
\end{equation}
where we have used $v_\chi\gg v_{\rho,\eta}$ for the case of $M_Z$
and $M_{Z'}$. Notice that the $Z'^0$ boson has a mass proportional
to $v_\chi$ and, like the charged bosons $V^+,U^{++}$, must be very
massive. In the present model we have
\begin{equation}
\frac{M^2_Z}{M^2_W}=\frac{1+4t^2}{1+3t^2}
\label{e20}
\end{equation}
where $t=g'/g\equiv\tan\theta$, and in order to obtain the usual
relation $\cos^2\theta_W M^2_Z=M^2_W$, with
$\cos^2\theta_W\!\approx\!0.78$, we must have $\theta\!\approx\!54^o$ i.e.,
$\tan^2\theta\!\approx\!11/6$. Then, we can identified $Z^0$ as the
neutral gauge boson of the standard model.

The neutral physical states are:
\begin{eqnarray}
A_\mu&=&\frac{1}{(1+4t^2)^{\frac{1}{2}}}\left[(W^3_\mu-
\sqrt3W^8_\mu)t+B_\mu\right],
\nonumber \\
Z^0_\mu&\simeq&-\frac{1}{(1+4t^2)^{\frac{1}{2}}}\left[(1+3t^2)
^{\frac{1}{2}}W^3_\mu+
\frac{\sqrt3t^2}{(1+3t^2)^{\frac{1}{2}}}W^8_\mu-\frac{t}{(1+3t^2)^
{\frac{1}{2}}}B_\mu\right], \nonumber \\
Z'^0_\mu&\simeq&\frac{1}{(1+3t^2)^{\frac{1}{2}}}\left(W^8_\mu+\sqrt3tB_
\mu\right).
\label{e21}
\end{eqnarray}
Con\-cer\-ning the vec\-tor bo\-sons, we ha\-ve the\-
fo\-llo\-wing tri\-li\-ne\-ar in\-te\-rac\-tions: $W^+W^-N$, $V^+V^-N$,
$U^{++}U^{--}N$ and $W^+V^+U^{--}$, where $N$ could be any of the
neu\-tral vec\-tor bo\-sons $A,Z^0$ or $Z'^0$.
\subsection{Charged and Neutral Currents}
\label{subsec:cc}
The interactions among the gauge bosons and fermions are read off from
\begin{equation}
{\cal L}_F=\bar Ri\gamma^\mu(\partial_\mu+ig'B_\mu N)R+
\bar Li\gamma^\mu(\partial_\mu+ig'B_\mu
N+\frac{ig}{2}\vec\lambda\cdot \vec W_\mu)L,
\label{e22}
\end{equation}
where $R$ represents any right-handed singlet and $L$ any left-handed
triplet.

Let us consider first the leptons. For the charged leptons, we have
the electromagnetic interaction by identifying the electron charge
as (see the Appendix)
\begin{equation}
e=\frac{g\sin\theta}{(1+3\sin^2\theta)^{\frac{1}{2}}}=\frac{g'\cos\theta}
{(1+3\sin^2\theta)^{\frac{1}{2}}},
\label{e23}
\end{equation}
and the charged current interactions are
\begin{equation}
{\cal L}_l^{CC}=-\frac{g}{\sqrt2}\sum_l\left(\bar\nu_{lL}\gamma^\mu
l_LW^+_\mu+ \bar l^c_L\gamma^\mu\nu_{lL} V^+_\mu+\bar l^c_L\gamma^\mu l_L
U^{++}_\mu+h.c.\right).
\label{e24}
\end{equation}
Notice that as we have not assigned to the gauge bosons a lepton
number, we have explicit breakdown of this quantum number
induced by the $V^+,U^{++}$ gauge bosons. A similar mechanism for
lepton number violation was proposed in Ref.~\cite{vs} but in that
reference the lepton-number-violating currents are coupled to the
standard gauge bosons and they are proportional to a small parameter
appearing in this model.

For the first generation of quarks we have the following charged
current interactions:
\begin{equation}
{\cal L}^{CC}_{Q_1W}=-\frac{g}{\sqrt2}\left(\bar u_L\gamma^\mu
d_{\theta L}W^+_\mu+\bar J_{1L}\gamma^\mu u_LV^+_\mu+\bar d_{\theta
L}\gamma^\mu J_{1L}U^{--}+h.c.\right),
\label{e25}
\end{equation}
and, for the second generation of quarks we have
\begin{equation}
{\cal L}^{CC}_{Q_2W}=-\frac{g}{\sqrt2}\left(\bar c_L\gamma^\mu
d_{\theta L}W^+_\mu-\bar s_{\theta L}\gamma^\mu J_{2\phi L}V^+_\mu+\bar
c_L\gamma^\mu  J_{2\phi L}U^{--}+h.c.\right).
\label{e26}
\end{equation}
The charge changing interactions for the third generation of quarks
are obtained from those of the second generation, making
$c\to t$, $s\to b$ and $J_2\to J_3$.
We have mixing only in the $Q\!\!=\!\!-\frac{1}{3}$ and
$Q\!\!=\!\!-\frac{4}{3}$ sectors, then in Eqs.~(\ref{e25}) and
(\ref{e26}) $d_\theta,s_\theta$ and $J_{2\phi}$ mean
Cabibbo-Kobayashi-Maskawa states in the three and two-dimensional
flavor space $d,s,b$ and $J_2,J_3$ respectively.
\subsection{Neutral Currents}
\label{sec:nc}
Similarly, we have the neutral currents coupled to both $Z^0$ and
$Z'^0$ massive vector bosons, according to the Lagrangian
\begin{equation}
{\cal L}_\nu^{NC}=-\frac{g}{2}\frac{M_Z}{M_W}
\bar\nu_{lL}\gamma^\mu\nu_{lL}
[Z_\mu-\frac{1}{\sqrt3}\frac{1}{\sqrt{h(t)}}Z'_\mu],
\label{e27}
\end{equation}
with $h(t)=1+4t^2$, for neutrinos and
\begin{equation}
{\cal L}_l^{NC} =-\frac{g}{4}\frac{M_Z}{M_W}[\bar
l\gamma^\mu(v_l+a_l\gamma^5)lZ_\mu+ \bar
l\gamma^\mu(v'_l+a'_l\gamma^5)lZ'_\mu],
\label{e28}
\end{equation}
for the charged leptons, where we have used $\bar l^c_L\gamma^\mu
l^c_L=-\bar l_R\gamma^\mu l_R$ and defined
\begin{equation}
\begin{array}{clccc}
v_l= & -1/h(t),&a_l=& 1,&(a) \\
v'_l=& -\sqrt{3/h(t)},
& a'_l=& v'_l/3.&(b)
\end{array}
\label{e29}
\end{equation}
With $t^2=11/6$, $v_l$ and $a_l$ have the same values of the
standard model.

As it was said before, the quark representations in Eqs.~(\ref{e3})
and~(\ref{e7}) are symmetry eigenstates, that is, they are related to
the mass eigenstates by Cabibbo-like angles. As we have one triplet
and two antitriplets, it should be expected  to exist flavor changing
neutral currents. Notwithstanding, as we shall show below, when we
calculate the neutral currents explicitly, we find that {\it all} of
them, for the same charge sector, have equal factors and the
GIM~\cite{q} cancellation is automatic in neutral currents coupled
to $Z^0$. Remind that in the standard electroweak
model, the GIM mechanism is a consequence of having each charge
sector the same coupling with $Z^0$, for example for the charge
$+2/3$ sector,
\begin{equation}
v^U_{SM}=1-\frac{8}{3}\sin^2\theta_W,\quad a^U_{SM}=-1.
\label{e30}
\end{equation}
The Lagrangian interaction among quarks
and the $Z^0$ is
\begin{equation}
{\cal L}_{ZQ}=-\frac{g}{4}\frac{M_Z}{M_W}\sum_i[\bar
\Psi_i\gamma^\mu(v^i+a^i\gamma^5)\Psi_i]Z_\mu,
\label{e31}
\end{equation}
where $i=u,c,t,d,s,b,J_1,J_2,J_3$; with
\begin{equation}
\begin{array}{rlrrc}
v^U=& (3+4t^2)/3h(t),&a^U=&-1,&(a)\\
v^D=&-(3+8t^2)/3h(t),& a^D=&1,&(b)\\
v^{J1}=&-20t^2/3h(t),& a^{J_1}=&0,&(c)\\
v^{J_2}=v^{J_3}= & 16t^2/3h(t),& a^{J_2}=a^{J_3}=&0,&(d)
\end{array}
\label{e32}
\end{equation}
$U$, and $D$ mean the charge $+2/3$ and $-1/3$ respectively, the same
for $J_{1,2,3}$. Notice that as it was said above, there is not
flavor changing neutral current coupled to the $Z^0$ field and that
the exotic quarks couple to $Z^0$ only through vector currents. It is
easy to verify that for the $Q= \frac{2}{3},-\frac{1}{3}$ sectors the
respective coefficients $v$ and $a$ also coincide with those of the
standard electroweak model if $t^2=11/6$, as required to maintain the
relation $\cos\theta_W M_Z=M_W$.

The same cancellation does not happen with the corresponding currents
coupled to the $Z'^0$ boson, each quark having its respective
coefficients. Explicitly, we have
\begin{equation}
{\cal L}_{Z'Q}=-\frac{g}{4}\frac{M_Z}{M_W}\sum_i[\bar
\Psi_i\gamma^\mu(v'^i+a'^i\gamma^5)\Psi_i]Z'_\mu,
\label{e33}
\end{equation}
where
\begin{equation}
\begin{array}{rlrlc}
v'^u=&-(1+8t^2)/\sqrt{3h(t)},&a'^u=
&1/\sqrt{3h(t)},&(a)\\
v'^c=v'^t=&(1-2t^2)/\sqrt{3h(t)},&
a'^c=a'^t=&-(1+6t^2)/\sqrt{3h(t)},&(b)\\
v'^d=&-(1+2t^2)/\sqrt{3h(t)},&a'^d=&-a'^c,&(c)\\
v'^s=v'^b=&\sqrt{h(t)/3}, & a'^s=a'^b=& -a'^u,&(d)
\end{array}
\label{e34}
\end{equation}
for the usual quarks, and
\begin{equation}
\begin{array}{rcrlc}
v'^{J_1}=&\frac{2}{\sqrt3}\frac{1-7t^2}{\sqrt{h(t)}},
&a'^{J_1}=&-\frac{2}{\sqrt3}\frac{1+3t^2}{\sqrt{h(t)}},&(a)\\
v'^{J_2}=v'^{J_3}=&-\frac{2}{\sqrt3}\frac{1-5t^2}{\sqrt{h(t)}},&
a'^{J_2}=a'^{J_3}=&-a'^{J_1},&(b)
\end{array}
\label{e35}
\end{equation}
for the exotic quarks.
\section{The Scalar Potential}
\label{sec:sp}
The most general gauge invariant potential involving the three Higgs
triplets is
\begin{eqnarray}
V(\eta,\rho,\chi)&=&\mu^2_1\eta^\dagger\eta+\mu^2_2\rho^\dagger\rho+
\mu^2_3\chi^\dagger\chi+\lambda_1(\eta^\dagger\eta)^2
+\lambda_2(\rho^\dagger\rho)^2+\lambda_3(\chi^\dagger\chi)^2
\nonumber \\ & &\mbox{}
+(\eta^\dagger\eta)[\lambda_4\rho^\dagger\rho+\lambda_5\chi^\dagger\chi]+
\lambda_6(\rho^\dagger\rho)(\chi^\dagger\chi)\nonumber \\ & &\mbox{}
+\sum_{ijk}\epsilon^{ijk}(f\eta_i\rho_j\chi_k+h.c.).
\label{36}
\end{eqnarray}
the coupling $f$ has dimesion of mass. We can analyse the scalar
spectrum defining
\begin{equation}
\eta^0=v_1+H_1+ih_1,\quad \rho^0=v_2+H_2+ih_2,\quad\chi^0=v_3+H_3+ih_3,
\label{e37}
\end{equation}
where we have redefined $v_\eta/\sqrt2,v_\rho/\sqrt2$ and
$v_\chi/\sqrt2$ as $v_1,v_2$ and $v_3$ respectively, and for
simplicity we are not considering relative phases between the vacuum
expectation values. Here we are only interested in the charged scalars
spectrum. Requiring that the shifted potential has no linear terms in
any of the $H_i$ and $h_i$ fields, $i=1,2,3$ we obtain in the tree
approximation the following constraint equations:
\begin{equation}
\begin{array}{r}
\mu^2_1+2\lambda_1v_1^2+\lambda_4v^2_2+\lambda_5v_3^2+Re\,fv_1^{-1}v_2v_3=0,\\
\mu^2_2+2\lambda_2v_2^2+\lambda_4v^2_1+\lambda_6v_3^2+Re\,fv_1v_2^{-1}v_3=0,\\
\mu^2_3+2\lambda_3v_3^2+\lambda_5v^2_1+\lambda_6v_2^2+Re\,fv_1v_2v_3^{-1}=0,\\
Im\,f=0.
\end{array}
\label{e38}
\end{equation}
Then, it is possible to verify that there is a doubly charged
Goldstone boson and a doubly charged physical scalar. There are also two
singly charged Goldstone bosons
\begin{equation}
\begin{array}{l}
G^-_1=(-v_1\eta^-_2+v_3\chi^-)/(v^2_1+v_3^2)^{\frac{1}{2}}, \\
G^-_2=(-v_1\eta^-_1+v_2\rho^-)/(v^2_1+v_2^2)^{\frac{1}{2}},
\end{array}
\label{e39}
\end{equation}
and two singly charged physical scalars
\begin{equation}
\begin{array}{c}
\phi^-=(v_3\eta^-_2+v_1\chi^-)/(v^2_1+v^2_3)^{\frac{1}{2}},\\
\varphi^-=(v_2\eta^-_1+v_1\rho^-)/(v^2_1+v^2_2)^{\frac{1}{2}},
\end{array}
\label{e40}
\end{equation}
with masses $m^2_1=fv_2(v_1^{-1}v_3+v_1v_3^{-1})$ and
$m_2^2=fv_3(v_1^{-1}v_2+v_2^{-1}v_1)$ respectively. We can see from
Eq.~(\ref{e40}) that the mixing occurs between $\eta_2^-$ and
$\chi^-$, $\eta_1^-$ and $\rho^-$ but not between $\eta^-_1$ and
$\eta^-_2$. This implies that the neutrinoless double beta decay does
not occur in the minimal model. It is necessary to introduce two new
Higgs triplets, say, $\sigma,\omega$ with the quantum numbers of
$\eta$ to have mixing between $\eta^-_1$ and $\eta^-_2$. In this case
the potential has terms with $\eta\to\sigma,\omega$ in
Eq.~(\ref{36}) and terms which mix $\eta,\sigma$ and $\omega$. In
particular the term $\epsilon^{ijk}\eta_i\sigma_j\omega_k$ mixs
$\eta^-_1,\sigma^-_1,\omega^-_1$ with
$\eta^-_2,\sigma^-_2,\omega^-_2$~\cite{pp}.
\section{Phenomenological Consequences}
\label{sec:pc}
In this model, the lepton number is violated in the heavy charged
vector bosons exchange but it is not in the neutral exchange ones,
because neutral interactions are diagonal in the lepton sector.
However, we have flavor changing neutral currents in the quark sector
coupled to the heavy neutral vector boson $Z'^0$. All these heavy bosons
have a mass which depends on $v_\chi$ and this vacuum expectation
value is, in principle, arbitrary.

Processes like $\mu^-\to e^-\nu_e\bar\nu_\mu$ are the typical
ones, involving leptons, which are induced by lepton-number-violating
charged currents in the present model. It is well known that the
ratio
\begin{equation}
R=\frac{\Gamma(\mu^-\to
e^-\nu_e\bar \nu_\mu)}{\Gamma(\mu^-\to all)}
\label{e41}
\end{equation}
tests the nature of the lepton family number conservation, i.e.,
additive vs. multiplicative. Roughly we have
\[R\propto \frac{A(3.a)}{A(3.b)}\approx
\left(\frac{M_W}{M_V}\right)^4\]
where $A(3.a)$ and $A(3.b)$ are the amplitudes for the processes in
Fig.\ \ref{fig3}(a) and (b) respectively.
Experimentally $R<5\times
10^{-2}$~\cite{pdg}, then we have that the occurrence of the decay
$\mu^-\to e^-\nu_e\bar\nu_\mu$ implies that $M_V>2M_W$.

In addition to decays, effects like $e^+_Le^-_R\to
\nu_{eL}\bar\nu_{eR}$ will also occur in accelerators, but
these events impose constraints on the masses of the vector bosons
which are weaker than those coming from the decays. Notice that the
incoming negative charged lepton is right-handed because the
lepton-number-violating interactions with the $V^+$ vector boson in
Eq.~(\ref{e24}) is a right-handed current for the electron.

The doubly charged vector boson $U^{--}$ will produce deviations from
the pure QED Moller scattering which could be detected at high
energies.

Stronger bounds on the masses of the exotic vector bosons come from
flavor changing neutral currents induced by $Z'^0$.
The contribution to the $K^0_L-K^0_S$ mass difference due to the
exchange of a heavy neutral boson $Z'^0$ appears in Fig.\ \ref{fig4}.
{}From Eq.~(\ref{e33}) we have explicitly
\begin{equation}
-\frac{g}{4}\frac{M_Z}{M_W}\cos\theta_c\sin\theta_c
[\bar d\gamma^\mu(v'^d+a'^d\gamma^5)s+\bar
d\gamma^\mu(v'^s+a'^s\gamma^5)s]Z^{0'}_\mu,
\label{e42}
\end{equation}
with $v'^{d,s}$ and $a'^{d,s}$ given in Eq.~(\ref{e34}c,d)
respectively, and for simplicity we have assumed only two-family
mixing. Then, Eq.~(\ref{e42}) produces at low energies
the effective interaction,
\begin{equation}
{\cal L}_{eff}=\frac{g^2}{16}\left(\frac{M_Z}{M_W}\right)^2
\frac{\cos^2\theta_c\sin^2\theta_c}
{M^2_{Z'^0}}\left[\bar d\gamma^\mu(c_v+c_a\gamma^5)s\right]^2,
\label{e43}
\end{equation}
where we have defined
\begin{equation}
\begin{array}{ccl}
c_v\equiv&v'^d-v'^s=&-\frac{2}{\sqrt3}(1+3t^2)/\sqrt{h(t)},\\
c_a\equiv&a'^d-a'^s=&-c_v.
\end{array}
\label{e44}
\end{equation}
The contribution of the c-quark in the standard model is~\cite{gl}:
\begin{equation}
{\cal L}^{SM}_{eff}=-\frac{G_F}{\sqrt2}\frac{\alpha}{4\pi}
\frac{m_c^2}{M_W^2\sin^2\theta_W}\cos^2\theta_c\sin^2\theta_c
[\bar d\gamma^\mu\frac{1}{2}(1-\gamma^5) s]^2,
\label{e45}
\end{equation}
with $g^2/8M_W^2=G_F/\sqrt2$. We can obtain the constraint upon the
neutral $Z'^0$ mass assuming, as usual, that any
additional contribution to the $K^0_S-K^0_L$ mass difference from the
$Z'^0$ boson cannot be much bigger than the contribution of the
charmed quark~\cite{ch}. Then, from Eqs.~(\ref{e43}) and~(\ref{e45})
we get
\begin{equation}
M^2_{Z'^0}>\left(\frac{1}{2}\frac{4\pi}{\alpha}c^2_a\frac{M^2_W}{m_c^2}
\tan^2\theta_W\right)M^2_W,
\label{e46}
\end{equation}
which implies the following lower bound on the mass of the $Z'^0$:
\[M_{Z'^0}>40\, TeV.\]
{}From this value and Eq.~(\ref{e19}) we see that  $v_\chi$ must
satisfy
\[v^2_\chi> \frac{3\sqrt2}{8G_FM_W^2(1+3t^2)}(40\,TeV)^2,\]
that is, $v_\chi>12 \,TeV$. As the vacuum expectation value
of the $\chi$ Higgs is $<\!\!\chi^0\!\!>=v_\chi/\sqrt2$ then we have that
$<\!\!\chi^0\!\!>\,>\!8.4\, TeV$. This also implies, from
Eq.~(\ref{e17}), that the masses of the charged vector bosons
$V^-,U^{--}$ are larger than $4\,TeV$.
\section{Conclusions}
\label{sec:conclusions}
If we admit lepton number violation, $SU(3)$ could be a good symmetry
at high energies, at least for the lightest leptons $(\nu,e^-,e^+)$.
Assuming that this is a local gauge symmetry, the rest of the model
follows naturally, including the exotic quarks, $J$'s.  To the
best of our knowledge, there is not laboratory or
cosmological/astrophysical constraints to the masses of the exotic
quarks (``Josions'')-- $J_1$ and $J_{2,3}$ with charge
$+\frac{5}{3}$ and $-\frac{4}{3}$ respectively but, they must be too
massive to be detected by present accelerators. For the case of the
heavy vector bosons, charged $U,V$ (``Wanios'')
and the neutral $Z'^0$ (``Zezeons''),
rare decays restraint their masses as we have
shown before. It is interesting to note that no extremely high mass
scale emerges in this model making possible its experimental test in
future accelerators.

Vertices like the following, appear in the scalar-vector sector:
\begin{equation}
\frac{ig}{\sqrt2}\left[W^+_\mu\left(\eta_1^-\partial^\mu\eta^0 -
\partial^\mu\eta_1^-\eta^0\right)+V^-_\mu\left(\eta_2^+\partial^\mu\eta^0-
\partial^\mu\eta_2^+\eta^0\right)\right],
\label{e47}
\end{equation}
and also with $\eta\to \sigma,\omega$, when these two new
triplets are added to the model. Then we have mixing in the scalar
sector which imply 1-loop contributions to the $(\beta\beta)_{0\nu}$
involving the vector bosons $V^-,U^{--}$ but these are less than
contributions at tree level through scalar exchange~\cite{pp}. On the
other hand, this model cannot produce processes like $K^-\to
\pi^+e^-\mu^-$ and $\tau^-\to l^+\pi^-\pi^-$ with $l=e,\mu$.

Notice that the definition of the charge conjugation transformation
we have used in this work, see the Appendix, has physical
consequences only in the Yukawa interactions and in the currents
coupled to the heavy charged gauge bosons where an opposite sign
appears with respect to the usual definition of that transformation.

\acknowledgments

We would like to thank the Con\-se\-lho Na\-cio\-nal de
De\-sen\-vol\-vi\-men\-to Cien\-t\'\i \-fi\-co
e Tec\-no\-l\'o\-gi\-co (CNPq) for full (F.P.) and partial (V.P.) financial
support, M.C. Tijero for reading the manuscript and, finally C.O.
Escobar, M. Guzzo and A.A. Natale for useful discussion.

\unletteredappendix{}

In this appendix we shall treat in more detail how it is possible to
get a Yukawa interactions from Eq.~(\ref{e11}).

In the present model we have put together in the same multiplet the
charged leptons and their respective charge conjugated field. That
is, both of them are considered as the two independent fermion
degrees of freedom. If we use the usual definition of the charge
conjugation transformation $\psi^c=C\bar\psi^T$,
$\overline{\psi^c}=-\psi^TC^{-1}$ the Yukawa couplings in
Eq.~(\ref{e11}) vanish, including the mass terms. This is a
consequence of the degrees of freedom we have chosen.
Notwithstanding, it is possible to define the charge conjugation
operation as follows:
\[\psi^c=\gamma^5C\bar\psi^T,\,\,\overline{\psi^c}=\psi^TC^{-1}\gamma^5.\]
This definition is consistent with quantum electrodynamics since its
only effect is to change the sign of the mass term in the Dirac
equation for the charge conjugated spinor $\psi^c$ with respect to
the mass term of the spinor $\psi$, and it is well known that the
sign of the mass term in the Dirac equation has no physical
meaning. With the negative sign, the upper components of the spinor
are the ``large'' ones, and with the positive sign, the large
components are the lower ones~\cite{t}.

Using this definition it is easy to verify that
$\overline{l^c_R}l^c_L=-\bar l_Rl_L$ instead of
$\overline{l^c_R}l^c_L=+\bar l_Rl_L$, which follows using the usual
definition of the charge conjugation transformation.

On the other hand, the definition of charge conjugation we have used
in this work, produces the same effect as the usual one in bilinear
terms as the vector interaction. Then, in the kinetic term and the
vector interaction with the photon, it is not possible to distinguish
both definitions. For example, the kinetic terms in the model are
\[\sum_l(\bar l_Li\!\!\not\!\partial
l_L+\overline{\psi^c}i\!\!\not\!\partial l^c_L),\]
with $l=e,\mu,\tau$ and it can be written as
\[\sum_l(\bar l_Li\!\!\not\!\partial l_L+\bar l_Ri\!\!\not\!\partial
l_R),\]
where the right-handed electron has been interpreted as the absence
of a left-handed positron with $(-E,-\vec p)$.

For charged leptons we have the electromagnetic interaction
\[-e(\bar l_L\gamma^\mu l_L-\overline{l^c_L}\gamma^\mu l^c_L)A_\mu,\]
and using $\overline{l^c_L}\gamma^\mu l^c_L\!\!=\!\!-\bar l_R\gamma^\mu l_R$
we obtain the usual vector interaction $-e\bar l\gamma^\mu lA_\mu$,
but on the other hand, in the charged currents we have
$\overline{\nu^c_{lR}}l_L\!=\!-\bar l_R\nu_{lL}$.

\figure{ (a) $e^-e^-\to W^-V^-$ process induced by
right-handed currents, $L$ and $R$ denote the handedness of the
current at the vertex, and $q$ is the momentum transfer.  (b) Diagram
for $W^-W^-\to e^-e^-$ with massive Majorana neutrinos, both
vertices are left-handed.\label{fig1}}
 \figure{Diagram for $W^-V^-\to
e^-e^-$ due to the existence of right-handed current (a) and doubly
charged gauge boson (b).\label{fig2}}
\figure{ (a) Lepton number conserving process. (b) Lepton number
violating process. \label{fig3}}
\figure{$Z'^0$ exchange contribution to the effective Lagrangian
for $K_S-K_L$ mixing.\label{fig4}}

\begin{references}
\bibitem{bk} B. Kayser, F. Gibrat-Debu and F. Perrier, The Physics of
Massive Neutrinos, World Scientific, 1989.
\bibitem{q} C. Quigg, Gauge Theories of the Strong, Weak, and
Electromagnetic Interactions, The Benjamin/Cummings, 1983.
\bibitem{v} M. Singer, J. W. F. Valle and J. Schechter, Phys.
Rev.{\bf D22}, 738(1980).
\bibitem{su} J. Schechter and Y. Ueda, Phys. Rev. {\bf D8}, 484(1973).
\bibitem{ls} P. Langacker and G. Segr\`e, Phys. Rev. Lett. {\bf 39}, 259(1977).
\bibitem{fm} H. Fritzsch and P. Minkowski, Phys. Lett. {\bf B63}, 99(1976).
\bibitem{lw} B. W. Lee and S. Weinberg, Phys. Rev. Lett. {\bf 38}, 1237(1977).
\bibitem{clt} J.~M.~Cornwall, D.~N.~Levin and G.~Tiktopoulos, Phys. Rev.
{\bf D10}, 1145 (1974).
\bibitem{vs} J.W.F. Valle and M. Singer, Phys. Rev. {\bf D28}, 540(1983).
\bibitem{pp} F. Pisano and V. Pleitez, Neutrinoless Double Beta
Decay With Massless Neutrinos, Preprint IFT-P.07/92, submited for
publication.
\bibitem{pdg} Particle Data Group, Phys. Lett. {\bf B239}, 1(1990).
\bibitem{gl} M.K. Gaillard and B.W. Lee, Phys. Rev. {\bf D10},
897(1974); R. Shrock and S.B. Treiman, Phy. Rev. {\bf D19},
2148(1979).
\bibitem{ch} R.N. Cahn and H. Harari, Nucl. Phys. {\bf B176},
135(1980).
\bibitem{t} J.~Tiomno, Nuovo Cimento {\bf 1}, 226(1955).
\end{references}
\end{document}